\documentclass{article}

\usepackage[T1]{fontenc}
\usepackage[utf8]{inputenc}
\usepackage[english]{babel}

\usepackage{amsmath}
\usepackage{amssymb}
\usepackage{amsthm}
\usepackage{mathtools}
\mathtoolsset{showonlyrefs,showmanualtags}
\usepackage{soul}

\usepackage{microtype}
\usepackage{graphicx}
\usepackage{booktabs}
\usepackage{url}
\usepackage{hyperref}
\usepackage{tikz}
\usepackage{float}

\newcommand\independent{\protect\mathpalette{\protect\independenT}{\perp}}
\def\independenT#1#2{\mathrel{\rlap{$#1#2$}\mkern2mu{#1#2}}}

\def\E{{\rm E{}}}

\usetikzlibrary{shapes,decorations,arrows,calc,arrows.meta,fit,positioning}
\tikzset{
    -Latex,auto,node distance =1 cm and 1 cm,semithick,
    state/.style ={ellipse, draw, minimum width = 0.7 cm},
    point/.style = {circle, draw, inner sep=0.04cm,fill,node contents={}},
    bidirected/.style={Latex-Latex,dashed},
    el/.style = {inner sep=2pt, align=left, sloped}
}
\usepackage{bbm}
\newcommand{\bbone}{\mathbbm{1}}
\usepackage{enumitem}
\setlist[enumerate]{itemsep=0.1ex}
\setlist[itemize]{itemsep=0.1ex}

\DeclarePairedDelimiter\ceil\lceil\rceil

\providecommand{\bbone}{\mathbf{1}}
\DeclarePairedDelimiterXPP\indicator[1]{\bbone}{\lbrack}{\rbrack}{}{#1}

\DeclarePairedDelimiterXPP\expf[1]{\exp}{\lparen}{\rparen}{}{#1}
\DeclarePairedDelimiterXPP\logf[1]{\log}{\lparen}{\rparen}{}{#1}
\DeclarePairedDelimiterXPP\maxf[1]{\max}{\lparen}{\rparen}{}{#1}
\DeclarePairedDelimiterXPP\minf[1]{\min}{\lparen}{\rparen}{}{#1}

\DeclarePairedDelimiterXPP\func[2]{#1}{\lparen}{\rparen}{}{#2}

\makeatletter
\newcommand*{\tran}{{\mathpalette\@tran{}}}
\newcommand*{\@tran}[2]{\raisebox{\depth}{$\m@th#1\intercal$}}
\makeatother

\DeclarePairedDelimiterXPP\tnorm[1]{}{\lVert}{\rVert_{1}}{}{#1}
\DeclarePairedDelimiterXPP\enorm[1]{}{\lVert}{\rVert_{2}}{}{#1}
\DeclarePairedDelimiterXPP\inorm[1]{}{\lVert}{\rVert_{\infty}}{}{#1}
\DeclarePairedDelimiterXPP\pnorm[2]{}{\lVert}{\rVert_{#1}}{}{#2}

\DeclarePairedDelimiterXPP\detf[1]{\det}{\lparen}{\rparen}{}{#1}

\DeclareMathOperator{\trsym}{tr}
\DeclarePairedDelimiterXPP\tr[1]{\trsym}{\lparen}{\rparen}{}{#1}

\DeclareMathOperator{\diagsym}{diag}
\DeclarePairedDelimiterXPP\diag[1]{\diagsym}{\lparen}{\rparen}{}{#1}

\DeclareMathOperator{\ranksym}{rank}
\DeclarePairedDelimiterXPP\rank[1]{\ranksym}{\lparen}{\rparen}{}{#1}

\DeclareMathOperator{\vectorizesym}{vec}
\DeclarePairedDelimiterXPP\vectorize[1]{\vectorizesym}{\lparen}{\rparen}{}{#1}

\let\Prsym\Pr
\let\Pr\relax
\DeclarePairedDelimiterXPP\Pr[1]{\Prsym}{\lparen}{\rparen}{}{%
	#1}

\DeclarePairedDelimiterXPP\Prsub[2]{\Prsym_{#1}}{\lparen}{\rparen}{}{%
	#2}

\DeclareMathOperator{\Esym}{E}

\DeclarePairedDelimiterXPP\Esub[2]{\Esym_{#1}}{\lbrack}{\rbrack}{}{%
	#2}

\DeclareMathOperator{\Varsym}{Var}
\DeclarePairedDelimiterXPP\Var[1]{\Varsym}{\lparen}{\rparen}{}{%
	#1}

\DeclarePairedDelimiterXPP\Varsub[2]{\Varsym_{#1}}{\lparen}{\rparen}{}{%
	#2}

\DeclarePairedDelimiterXPP\EstVar[1]{\widehat{\Varsym}}{\lparen}{\rparen}{}{%
	#1}

\DeclareMathOperator{\Covsym}{Cov}
\DeclarePairedDelimiterXPP\Cov[1]{\Covsym}{\lparen}{\rparen}{}{%
	#1}

\DeclarePairedDelimiterXPP\Covsub[2]{\Covsym_{#1}}{\lparen}{\rparen}{}{%
	#2}

\DeclareMathOperator{\Corrsym}{Corr}
\DeclarePairedDelimiterXPP\Corr[1]{\Corrsym}{\lparen}{\rparen}{}{%
	#1}

\makeatletter
\newcommand{\indep}{\protect\mathpalette{\protect\@indep}{\perp}}
\newcommand*{\@indep}[2]{\mathrel{\rlap{$#1#2$}\mkern3mu{#1#2}}}
\makeatother

\newcommand{\bigOsym}{\mathcal{O}}
\DeclarePairedDelimiterXPP\bigO[1]{\bigOsym}{\lparen}{\rparen}{}{#1}

\newcommand{\littleOsym}{o}
\DeclarePairedDelimiterXPP\littleO[1]{\littleOsym}{\lparen}{\rparen}{}{#1}

\newcommand{\bigOpsym}{\bigOsym_p}
\DeclarePairedDelimiterXPP\bigOp[1]{\bigOpsym}{\lparen}{\rparen}{}{#1}

\newcommand{\littleOpsym}{\littleOsym_p}
\DeclarePairedDelimiterXPP\littleOp[1]{\littleOpsym}{\lparen}{\rparen}{}{#1}

\newcommand{\bigOmegasym}{\Omega}
\DeclarePairedDelimiterXPP\bigOmega[1]{\bigOmegasym}{\lparen}{\rparen}{}{#1}

\newcommand{\littleOmegasym}{\omega}
\DeclarePairedDelimiterXPP\littleOmega[1]{\littleOmegasym}{\lparen}{\rparen}{}{#1}

\newcommand{\bigThetasym}{\Theta}
\DeclarePairedDelimiterXPP\bigTheta[1]{\bigThetasym}{\lparen}{\rparen}{}{#1}

\theoremstyle{plain}

\newtheorem{proposition}{Proposition}

\theoremstyle{definition}

\theoremstyle{remark}

\usepackage[accepted]{icml2021}

\icmltitlerunning{Identification is not enough, but RCTs are}

\begin{document}

\twocolumn[
\icmltitle{Nonparametric identification is not enough, but randomized controlled trials are}

\icmlsetsymbol{equal}{*}

\begin{icmlauthorlist}
\icmlauthor{P. M. Aronow}{yale}
\icmlauthor{James M. Robins}{hsph}
\icmlauthor{Theo Saarinen}{ucb}
\icmlauthor{Fredrik S\"avje}{yale}
\icmlauthor{Jasjeet Sekhon}{yale}

\end{icmlauthorlist}

\icmlaffiliation{yale}{Yale University}
\icmlaffiliation{ucb}{University of California, Berkeley}
\icmlaffiliation{hsph}{Harvard School of Public Health}

\icmlcorrespondingauthor{P. M.  Aronow}{peter.aronow@yale.edu}

\icmlkeywords{Keyword 1, Keyword 2}

\vskip 0.3in
]

\printAffiliationsAndNotice{}

\begin{abstract}

We argue that randomized controlled trials (RCTs) are special even among settings where average treatment effects are identified by a nonparametric unconfoundedness assumption. This claim follows from two results of Robins and Ritov (1997): (1) with at least one continuous covariate control, no estimator of the average treatment effect exists which is uniformly consistent without further assumptions, (2) knowledge of the propensity score yields a uniformly consistent estimator and honest confidence intervals that shrink at parametric rates with increasing sample size, regardless of how complicated the propensity score function is.  We emphasize the latter point, and note that successfully-conducted RCTs provide knowledge of the propensity score to the researcher. We discuss modern developments in covariate adjustment for RCTs, noting that statistical models and machine learning methods can be used to improve efficiency while preserving finite sample unbiasedness. We conclude that statistical inference has the potential to be fundamentally more difficult in observational settings than it is in RCTs, even when all confounders are measured.
\end{abstract}
{\it One of my goals in this chapter is to explain, from the point of view of causal diagrams, precisely why RCTs allow us to estimate the causal effect $X \rightarrow Y$ without falling prey to confounder bias. Once we have understood why RCTs work, there is no need to put them on a pedestal and treat them as the gold standard of causal analysis, which all other methods should emulate. 

$\qquad\qquad\qquad\quad$ - Judea Pearl,  \citep{pearl2018book}}

\section{Introduction}

Identification of a parameter under a given model means that we would be able to calculate that parameter's value if the entire population distribution were observed. A common source of identification is an assumption of unconfoundedness, as could be motivated by the following causal graph including nodes for a treatment $X$, outcome $Y$, and covariate $W$.

\begin{figure}[h]
\begin{center}
\begin{tikzpicture}
    \node[state] (d) at (0,0) {$X$};

    \node[state] (x) at (1,1) {$W$};

    \node[state] (y) at (2,0) {$Y$};


    \path (x) edge (y);
    \path (d) edge (y);
    \path (x) edge (d);

\end{tikzpicture}
\end{center}
\caption{A Directed Acyclic Graph Satisfying Unconfounded Treatment Assignment}
\end{figure}
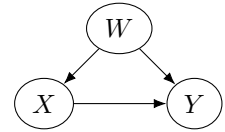

Assuming the graphical structure in Figure 1, the average treatment effect of $X$ on $Y$ is identified by the back-door criterion under \citet{pearlbook}'s causal model, and can be expressed as:
\begin{equation}
    \E \big[ \E[Y \vert X = 1, W] - \E[Y \vert X = 0, W]\big],
\end{equation} which involves only observable population characteristics. Therefore, under this model, the average treatment effect is nonparametrically identified.

However, identification of this sort is not enough if researchers seek to draw conclusions from actual data, no matter the number of observations they have access to. We argue that for inquiry into average potential outcomes (and therefore average treatment effects), successfully-conducted RCTs are special, even among the class of studies where treatment is unconfounded. That is, we show that the benefits of an RCT go beyond ensuring that unconfoundedness holds, and also include statistical guarantees for estimation and inference. Our claims are motivated by two results from a foundational article by \citet{Robins1997Curse} (henceforth Robins-Ritov).

We begin by revisiting a key negative finding from Robins-Ritov about observational studies: nonparametric identification of the sort obtained by Figure 1 is not sufficient for researchers to obtain accurate estimates. Even when the sample is large and treatment assignment is ignorable (or ``as-if randomly assigned conditional on covariates"), there remains a difficult estimation challenge that can only be resolved by adding additional, unverifiable assumptions.\footnote{Assumptions often used in the literature to facilitate estimation include: smoothness restrictions in the outcome and/or propensity score models \citep{robins2008higher,Armstrong_2018,kennedy2020optimal}, sparsity assumptions when covariates are explicitly high-dimensional \citep{Farrell_2015,kennedy2020optimal}, assumed rates on the estimation error for nuisance parameters \citep{chernozhukov2017doubledebiased}, and shape constraints on the outcome model \citep{Cai2013,Armstrong_2018}.} This argument, to our knowledge, was first made rigorously by Robins-Ritov, with discussions by \citet{wasserman1,wasserman2}; see also \citet{ritovbickel}. These results imply that nonparametric identification of a causal functional, be it from a model of potential outcomes, or from a graphical model, is not enough to guarantee that any uniformly consistent estimator is available.

We then demonstrate that correctly implemented RCTs offer a guarantee that unconfoundedness alone does not: they guarantee the validity of inferential procedures without requiring further assumptions. Building on known results on the Horvitz-Thompson \citep{HorvitzThompson} estimator, Robins-Ritov demonstrated that knowledge of the propensity score is sufficient to guarantee that there exist provably good (e.g., root-$n$ consistent, unbiased) estimators and inferential procedures. Since RCTs guarantee researcher's knowledge of the assignment mechanism (and therefore propensity scores), challenges with estimation and inference are obviated once weak regularity conditions are satisfied. Thus we argue that RCTs are special in the class of nonparametrically identified studies, because RCTs facilitate estimation and inference without the need for additional assumptions about the complexity of the regression problem. 





\section{Estimation and inference in observational studies and RCTs}

We begin our discussion by adapting the example presented by Robins and Wasserman \citep{wasserman1} regarding the Robins-Ritov example, with our exposition largely following theirs.

We restrict our attention to assignment mechanisms that are ``regular,'' as defined by \citet{Imbens2015Causal}, which rules out certain mechanisms that induce dependence between treatments.
In practice, we assume that we have an i.i.d.\ sample from $(W_i,Y_i,X_i)$, where $Y_i$ and $X_i$ are binary, and $W_i$ is drawn from a continuous distribution (either known or unknown). The assumption that $W_i$ is continuous is key to our discussion: it implies that we will never observe two units with the same $W_i$ values. As Robins-Ritov note, this is one manifestation of a ``curse of dimensionality" problem, and the argument qualitatively may hold with high-dimensional discrete data.

For our discussion, we can invoke the consistency assumption (c.f., the stable-unit-treatment-value-assumption), $$Y_i = Y_i(0)(1-X_i) + Y_i(1) X_i,$$ and unconfoundedness, $$ Y_i(x) \independent X_i | W_i, \forall x \in \{0,1\}.$$ This unconfoundedness assumption could be motivated by invoking the backdoor criterion on a graphical causal model such as the one depicted in Figure 1. For the purposes of this paper, we set aside concerns about the generic plausibility of this assumption in observational settings. We further define the propensity score, $\pi(W_i) = \Pr{X_i = 1 | W_i}$, where it is assumed that positivity holds: $$\delta < \pi(W_i) < 1-\delta$$ with probability one for some known positive $\delta$ bounded away from 0. 

In an RCT, all determinants of treatment are known because the experimenter decides on the assignment.\footnote{We continue to maintain the assumption that the sample is i.i.d.\ to ensure comparability with the observational setting. However, this is a surprisingly restrictive assumption in practice, as it implies that sample observations are independent, which rarely is the case in RCTs. A more general approach considers the assignment and sampling processes separately. See \citet{Abadie2020Sampling} and \citet{savje} for discussions related to this point.} (We assume full compliance with the treatment assignment and complete ascertainment of the outcome $Y_i$.) Thus, the backdoor criterion again can be invoked. Similarly, positivity is guaranteed by the design. The fact that RCTs can be designed to ensure that these conditions hold is well-known. Our argument rests on one perhaps less-appreciated fact: in an RCT, the propensity score, $\pi(W_i)$, is also known to the researcher, since it is specified by the experimental design. 

Without loss of generality, our interest is in learning the average potential outcome under treatment $\mu_1 = \E[Y_i(1)].$ A key result of Robins-Ritov (Theorem 3) implies the following:

\begin{proposition}
\label{prop:proposition_1}
Suppose we have a study where the consistency assumption, unconfoundedness, and positivity hold. Without further assumptions, no estimator of $\mu_1$ can be guaranteed to be uniformly consistent.
\end{proposition}

 The Robins-Ritov proof entails constructing an adversarially defined distribution at each $n$, with increasingly small bins as the sample size grows. In particular, no bin includes two observations with high probability, and therefore the within-bin correlation between propensity scores and outcomes cannot be estimated from the data. In our simulations, we will reconstruct a setting approximating the Robins-Ritov asymptotics to illustrate the logic.  

Although the Robins-Ritov result is a negative one, scholars have been investigating what in fact can be learned under different types of complexity-reducing assumptions about the outcome regression and propensity score model. The bounds of what can be learned under different types of conditions are still being explored. Notable work at the frontier includes \citet{robins2008higher, robins2009, armstrong2021finite}. All of these highlight a fundamental trade-off: because we are in a world in which the assignment mechanism is unknown, the strength of our statistical procedures critically depend on the assumptions brought to bear.
In another strand of research, \citet{Maclaren2020What} provides an exposition of the estimation challenges associated with nonparametric identification by borrowing from multiple fields and building from first principles in category theory.

These result can be contrasted to a much more optimistic set of results for RCTs. To reiterate, in an RCT, we can ensure by design: (i) unconfoundedness conditional on known covariates $W_i$, (ii) positivity, and (iii) knowledge of the propensity score $\pi(W_i)$. Importantly, knowledge of the propensity score guarantees the existence of an unbiased estimator via the Horvitz-Thompson principle.\footnote{Robins-Ritov note that this result demonstrates the failure of likelihood principle in addressing infinite-dimensional problems. For related discussions, see \citet{godambe,basu,freedmaninfinite}.}

\begin{proposition}

Suppose we have an RCT. Then the Horvitz-Thompson estimator $\hat\mu_1 = n^{-1}\sum_{i=1}^n \frac{Y_i X_i}{\pi_i(W_i)}$ is unbiased, uniformly root-$n$ consistent, and asymptotically normal for $\mu_1$. 
\end{proposition}

The result follows directly from the law of iterated expectations, along with standard results for the sample mean. Finally, Robins-Ritov present a result that establishes a non-trivial confidence interval that is applicable when the propensity score is known. 

\begin{proposition}

Suppose we have an RCT. Then honest, finite-sample-valid $1-\alpha$ confidence intervals of width $O(n^{-1/2})$ for $\mu_1$ can be constructed as $\hat \mu_1 \pm \sqrt{(2n\delta^2)^{-1} \log{(2 \alpha^{-1})}}$.
\end{proposition}

A proof, as noted by \citet{wasserman1}, follows directly from Hoeffding's inequality. This confidence interval procedure is conservative relative to, e.g., normal-approximation-based approaches, but has the benefit of being finite-sample exact and therefore guaranteeing uniform validity. Importantly, these procedures -- for both estimation and inference -- can be refined by appeal to asymptotic properties. 

Building on foundational results from \citet{robinsrotnitzky},  \citet{yangtsiatis} highlighted that in randomized trials, statistical models can be used to adjust for covariates without requiring the validity of these models. Importantly, if used suitably, the adjusted estimator will remain finite-sample unbiased \citep{aronowmiddleton} even if the model is misspecified. Furthermore, if used suitably, these adjustments cannot harm and can dramatically increase asymptotic precision. 

To ground ideas, consider the following cross-fitting, regression-based estimator of $\mu_{1}$,

\begin{align}
 \hat \mu_{1,R} =  n^{-1} \sum_{i=1}^n \left\{ \frac{(Y_i -  g(W_i, \hat \beta_i))X_i}{\pi_i(W_i)} + \sum_{i=1}^n g(W_i, \hat \beta_i) \right\},
\end{align}
where $\hat \beta_i$ is a finite-dimensional parameter vector fit using the first $K=\ceil{N/2}$ observations if $i>K$, and $\hat \beta_i$ is fit using the last $N-K$ observations if $i\leq K$, and $g$ is a fixed, smooth function with codomain in $[0,1]$.
This is merely one type of estimator that could be chosen in this setting, and estimators that involve estimating a propensity score model that is known to nest truth may increase efficiency under further regularity conditions \citep[c.f.][]{robinsrotnitzky,hirano}


\begin{proposition}
\label{prop:prop_4}
Suppose that we have an RCT. Then (a) $\hat \mu_{1,R}$ is finite-sample unbiased for $\mu_{1}$. (b) If in addition $\hat \beta_{i}$ converges to $\beta$ in $L_2$
for {\bf any} $\beta$, then $\hat \mu_{1,R}$ is root-$n$ consistent. (c) If $\hat \beta_i$ is estimated by minimizing the empirical weighted mean square error of the outcome regression with weights $w_i = \frac{1-\pi(W_i)}{\pi(W_i)^2}$, then $\hat \mu_{1,R}$
is at least as asymptotically efficient as $\hat \mu_{1}$.
\end{proposition}

The supporting information contains a proof for Proposition 4(a). Proposition 4(b) follows from \citet{liu2020nearly}, Theorem 1, and Proposition 4(c) follows from results in \citet[sec.\ 3]{cao2009}. 

We note that in (b), we do not require that $\hat \beta_i$ converge to any correctly specified $\beta$, only to some $\beta$, and in (c), as the analyst knows the true propensity scores, the condition can be fulfilled using the exact weights required when estimating $\hat \beta_i$.%
\footnote{While our results were predicated on an i.i.d. sample, analogous results hold for design-based inference for experiments performed on finite populations. \citep[c.f.,][]{lin,aronowmiddleton,sekhonlasso,loop, middletonunified}}

\section{Simulation Study}

In order to showcase the practical benefit of knowledge of the true assignment mechanism, we present an array of simulations where we compare several estimators of the Average Treatment Effect (ATE) on simulated data in which the true assignment probabilities are known.

This simulated data allows us to consider estimators which make use of the true assignment probabilities, estimators which make use of estimated assignment probabilities, and estimators which make no use of assignment probabilities. 

\subsection{Simulated Data}
\label{sim_setups}

Below we outline the three simulated data setups we use to measure the performance of different estimators. 
In order to compare the performance of estimators which utilize the propensity score with those which do not, we simulate the treatment assignments and potential outcomes according to linear, smooth nonlinear, and extremely nonlinear functions.

Following the spirit of the example in \citet{Robins1997Curse}, we generate an i.i.d. sample from $(W_i,Y_i,X_i)$, where $Y_i$ and $X_i$ are binary, and $W_i \sim U(0,1)$.
For each different experiment, we use the same function to simulate both $P(X_i  = 1| W_i)$ and $P(Y_i = 1 | W_i)$ under the assumed sharp null hypothesis that $Y_i(1) = Y_i(0) = Y_i$.

For Experiment 1, we take the logistic regression model:
\begin{align*}
    P(Y_i = 1 | W_i) = P(X_i = 1 | W_i) = f(.5 W_i + .1),
\end{align*}
where
\begin{align*}
    f(x) = \frac{e^x}{1+e^x}
\end{align*}
For Experiment 2, we take the smooth sinusoidal function:
\begin{multline*}
    P(Y_i = 1 | W_i) = P(X_i = 1 | W_i)
    \\
    = .2 \sin(15W_i)+.4W_i + .1
\end{multline*}
For Experiment 3, we break the unit interval into $B = 100n$ consecutive bins of equal length. The number of bins grows as we increase $n$. 
Then, for $W_i \in B_j$ (where $B_j$ denotes the $j$th bin), we use a highly nonlinear function:
\begin{align*}
    P(Y_i = 1 | W_i) = P(X_i = 1 | W_i) = 
    \begin{cases} 
      .1 & \text{if } j \text{ mod } 2 = 0 \\
      .9 & \text{if } j \text{ mod } 2 = 1
   \end{cases}
\end{align*}


%
%
%

\subsection{Estimators}

In order to explore the difference in performance between estimators which make use of the known propensity score, and estimators which do not, we select a variety of estimators from each category.\footnote{Details of our implementation, and replication instructions can be found at \url{https://github.com/theo-s/identification-not-enough}.}


Specifically we use:
\begin{itemize}
    \item Linear logistic regression.
    \item Nearest Neighbor matching, as implemented in the \textbf{Matching} package \citep{MatchingPackage}.
    \item Random Forest, as implemented in the \textbf{causalToolbox} package \cite{causaltoolboxPackage}.
        \item Propensity Score Matching, as implemented in the \textbf{Matching} package \citep{MatchingPackage}, using the true propensity score.
    \item The Horvitz-Thompson estimator \citep{HorvitzThompson}. 
    \item The leave-one-out Random Forest-adjusted Horvitz-Thompson estimator, \citep{aronowmiddleton,loop} using random forest as implemented in the \textbf{Rforestry} package \cite{kunzel2021linear}.
    \item The cross-fitting Horvitz-Thompson estimator as studied in Proposition \ref{prop:prop_4}, using random forest as implemented in the \textbf{ranger} package \cite{wright2015ranger}.
    \item The Adjusted Horvitz-Thompson estimator using random forest for covariate adjustment as implemented in the \textbf{ranger} package \cite{wright2015ranger}.
    \item Double Robust estimator using linear logistic regression for both the outcome and propensity score estimation \cite{robins1994}.
    \item Double Robust estimator using random forest as implemented in the \textbf{ranger} package \cite{wright2015ranger} for both the outcome and propensity score estimation.
    \item Double Robust estimator using random forest as implemented in the \textbf{ranger} package \cite{wright2015ranger} for both outcome and propensity score estimation and cross fitting over three disjoint folds.
\end{itemize}

%
%
%
%
%
%
%
%
%

\subsection{Results}

For each of the three experimental setups, we benchmark the performance of each estimator across a range of sample sizes, with each point combination repeated over 500 Monte Carlo replications.

In Figure \ref{fig:rmse_sim_figure}, we display the Root Mean Square Error (RMSE) for each estimator in the three experiments.

In order to specify the category each estimator falls into, we use a solid line to specify estimators which use the true propensity score, a dashed line to specify the estimators which use estimated propensity scores, and a dotted line to specify the estimators which do not use the propensity score. 

\begin{figure}[H]
\centering
    \includegraphics[width=0.45\textwidth]{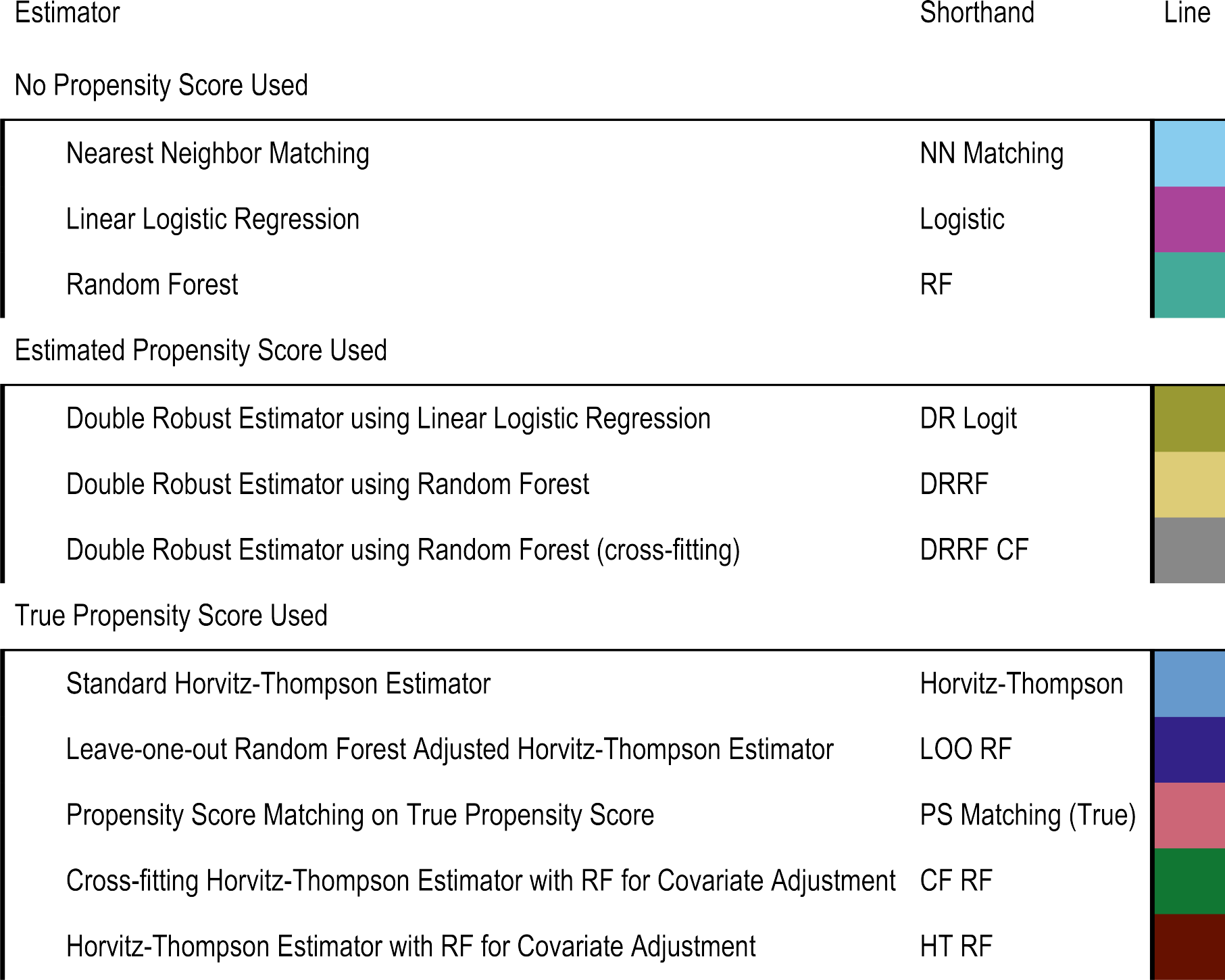}
    \label{fig:estimators_shorthand}
\end{figure}

\begin{figure}[H]
    \centering
    \includegraphics[width=0.15\textwidth]{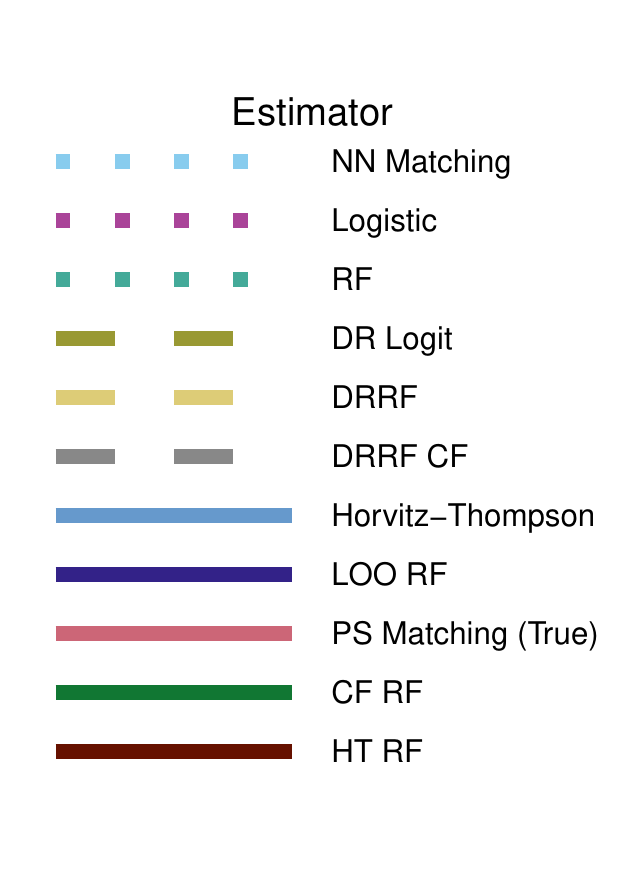}
    \label{fig:estimators_legend}
\end{figure}

\newpage


\begin{figure}[H]
    \centering
    \includegraphics[width=0.39\textwidth]{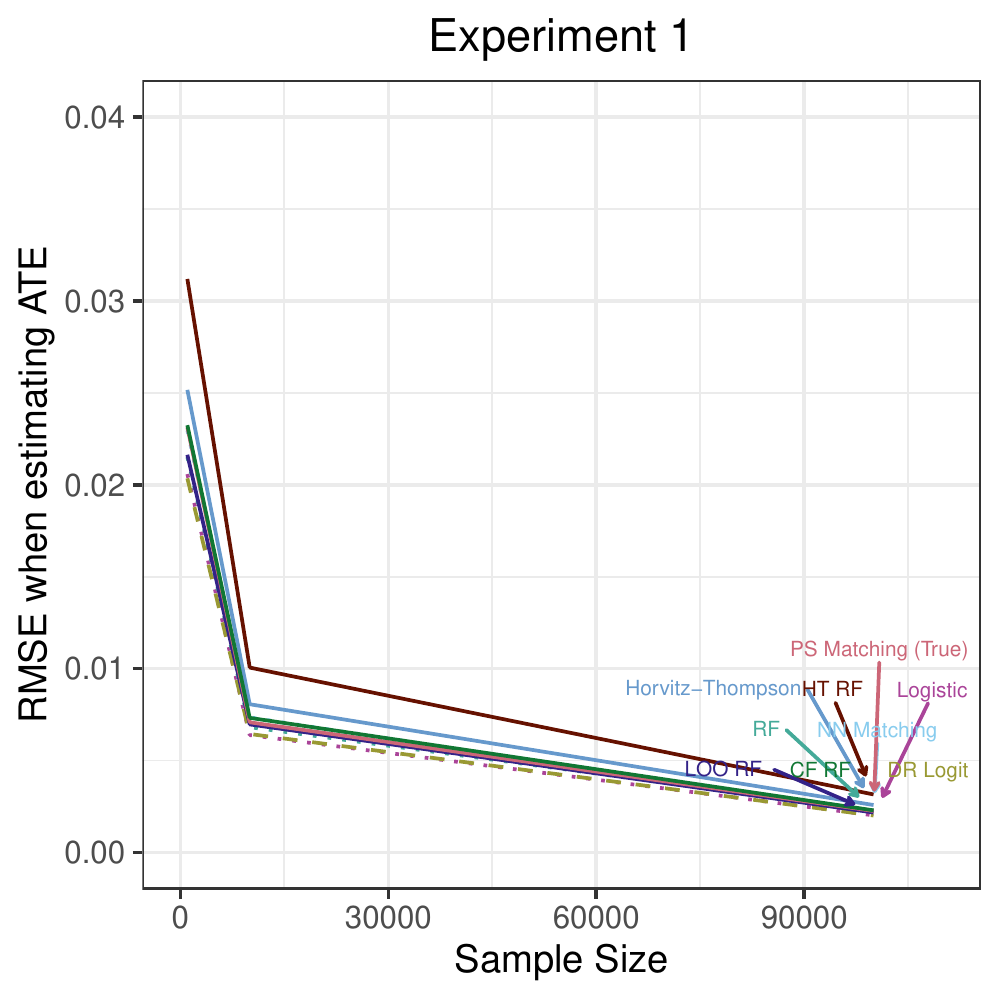}
    \includegraphics[width=0.39\textwidth]{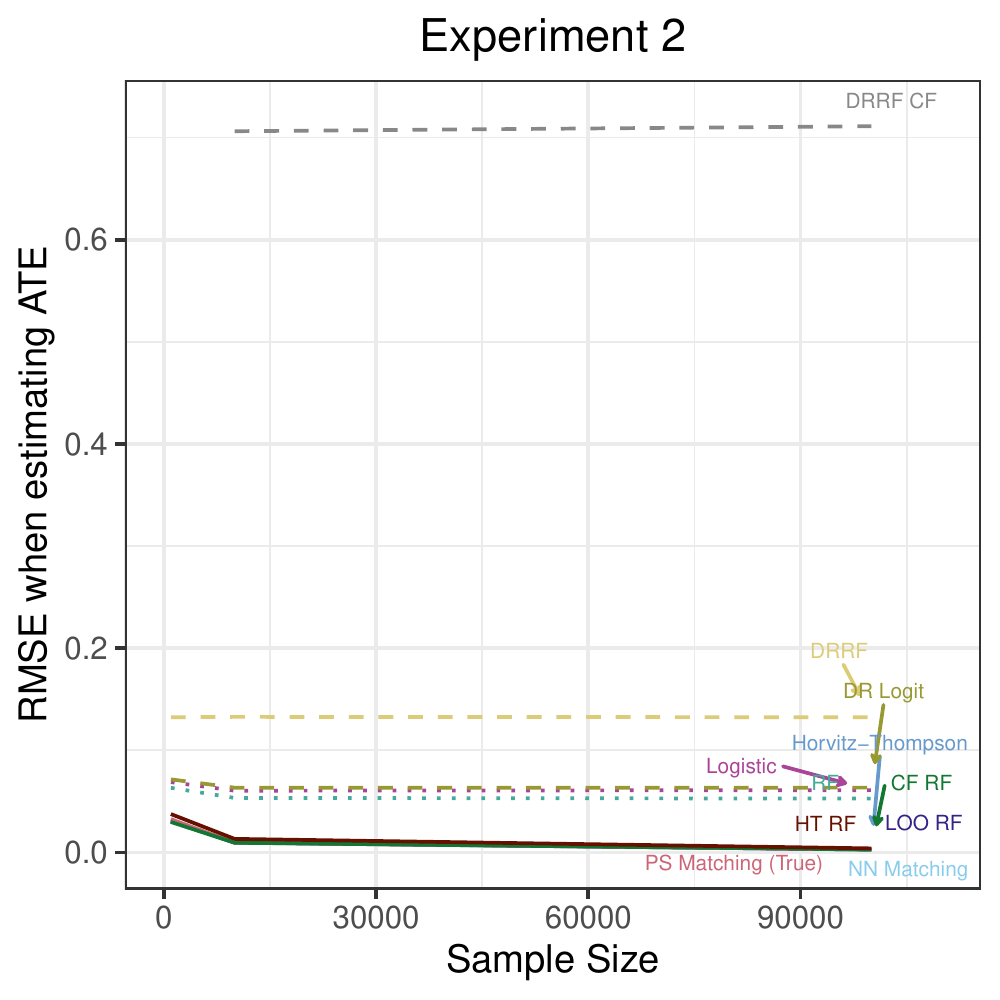}
    \includegraphics[width=0.39\textwidth]{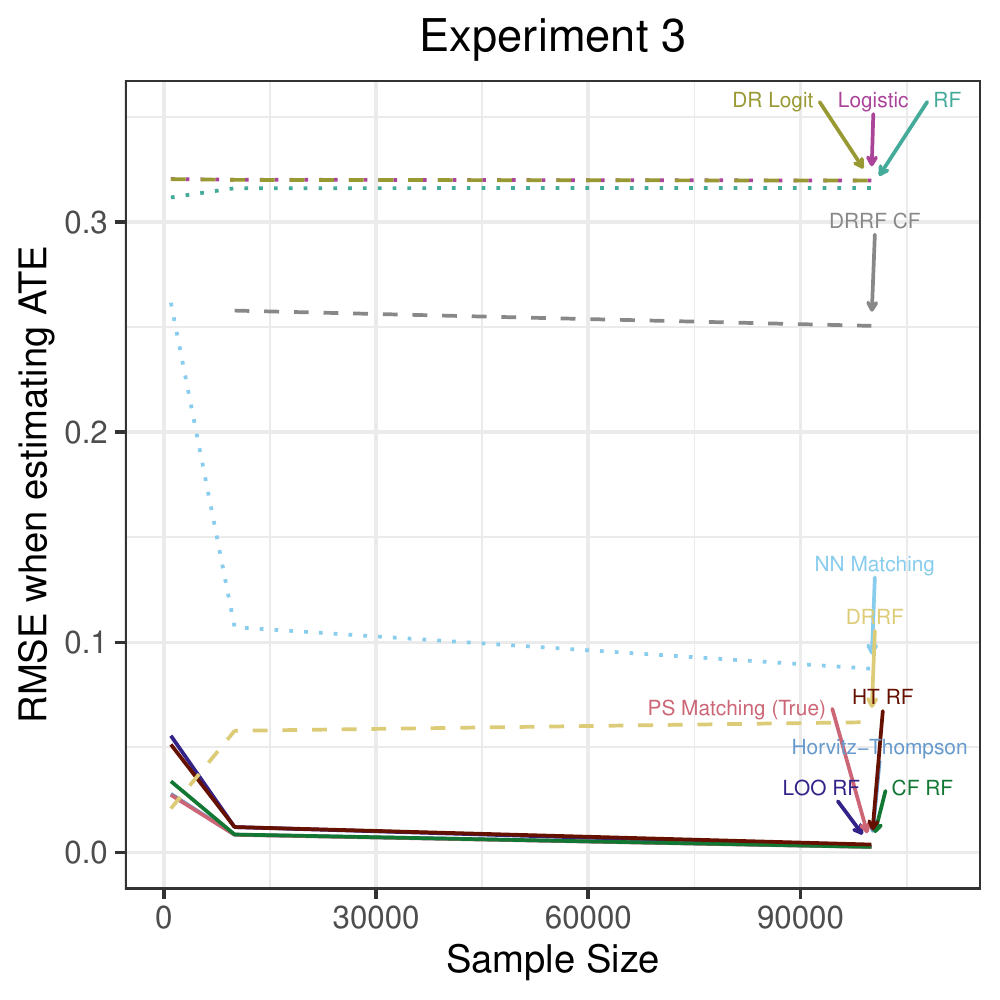}
    \caption{RMSE of several estimators when estimating the average treatment effect for a range of $n$ in the experimental setups outlined in Section \ref{sim_setups}}
    \label{fig:rmse_sim_figure}
\end{figure}

We see that the estimators which utilize the propensity score have significantly lower RMSE across all sample sizes.

The results of the Double Robust learner using random forests for the outcome and propensity score models are poor for the first two experiments, where the random forest seems to have trouble well-approximating the regression functions for the propensity score and potential outcomes. The RMSEs of the Double Robust learners were orders of magnitude higher in Experiment 1, so we have omitted their points from the first plot for clarity.

The first two experimental setups show the performance of estimators in settings in which the outcome and propensity score functions exhibit smoothness.
The third experiment borrows from from the original proof of Proposition \ref{prop:proposition_1} given in \citet{Robins1997Curse}. 
In the proof, Robins and Ritov consider a class of highly nonlinear laws in which consecutive dyadic intervals are given substantially different probability masses. 
In Experiment 3, we see that creating only a moderate number of such intervals exacerbates the performance of the estimators which do not utilize the propensity scores.
We note that methods that use the true propensity score outperform methods that use an estimated score in some of our simulations. 
We attribute this to bias in the estimated propensity score due to the adversarial nature of the data generating process and our choice of estimation algorithm.

\section{Discussion}

In our opening quote, Pearl asserted that RCTs do not provide a gold standard, and implies that RCTs deserve no special position relative to other studies that are nonparametrically identified by a causal diagram. However, insofar as a causal analysis requires working with actual data, we disagree with Pearl's conclusion, because his reasoning omits the sometimes-provably irresolvable challenge of statistical estimation and inference from observational studies. Even when a causal effect is nonparametrically identified, it still may not be possible to estimate it well. Even among unconfounded studies, RCTs are special because they can allow for estimation and inference at parametric rates under weak regularity conditions.

Although we have demonstrated that RCTs are special, they are not unique in providing knowledge of the propensity score. One key example relates to some {\it natural experiments}, though  nuance is required.\footnote{See \citet{titiunik2021natural} for a discussion of taxonomy relating to randomization.} Consider, for example, natural experiments that that arise in industry using recommender and other algorithmic systems that stochastically assign treatments. In such applications, the inputs of the algorithmic system are often known. However, the system's (probabilistic) treatment assignment function is generally not recorded and hence not available to the analyst. In cases where assignment probabilities cannot be explicitly recovered, the analyst typically attempts to adjust for the inputs to the algorithmic system. However, as our results demonstrate, such adjustments can perform poorly, even if identification holds. These estimation difficulties can be profound in algorithmic settings because the input space is often high-dimensional, and the treatment assignment function and the outcome regression tend be highly nonlinear, where slightly different values of the covariates result in substantially different probability masses. Therefore, it is worth ensuring that details about the assignment mechanism are recorded so as to facilitate estimation and inference at the analysis stage. While it might be infeasible to record the whole assignment function in these cases, it is  sufficient to record the assignment probabilities for the units in the sample.

The purpose of algorithmic assignment is often to use a Machine Learning (ML) system to assign treatments using complex functions of high-dimensional data at scale. People turn to ML systems because, in this setting, human analysts building parametric models of treatment assignment may perform poorly. However, this is precisely a setting in which estimation of the response or propensity score models may be difficult. To put it briefly: there is rarely a free lunch in statistics, but if one has knowledge of the propensity score, using it to avoid estimation difficulties can be one of them.



\section*{Acknowledgements} The authors thank Jonathon Baron, Victor Chernozhukov, Austin Jang, Joel Middleton, and Andrea Rotnitzky for helpful comments.

\bibliography{references}
\bibliographystyle{icml2021}

\newpage

\newpage
\appendix
\theoremstyle{plain}
\newtheorem{thm}{Theorem}
\newtheorem{prop}[thm]{Proposition}
\newtheorem{defn}[thm]{Definition}
\newtheorem{ex}[thm]{Example}

\makeatletter
\renewenvironment{proof}[1][\proofname] {\par\pushQED{\qed}\normalfont\topsep6\p@\@plus6\p@\relax\trivlist\item[\hskip\labelsep\bfseries#1\@addpunct{:}]\ignorespaces}{\popQED\endtrivlist\@endpefalse}
\makeatother

\definecolor{direct}{HTML}{FF0000}
\definecolor{indirect}{HTML}{FF9999}
\definecolor{dirin}{HTML}{990000}
\definecolor{control}{HTML}{999999}
\definecolor{directE}{HTML}{ED1C24}
\definecolor{indirectE}{HTML}{F69679}
\definecolor{controlE}{HTML}{999999}

\def\E{{\rm E}}
\def\V{{\rm V}}
\def\Cov{{\rm Cov}}
\def\I{{\rm I}}
\def\Supp{{\rm Supp}}
\def\Pr{P}
\def\Supphat{\text{S}\widehat{\text{up}}\text{p}}

\def\S{{\sigma}}
\newcommand{\X}{\mathbf{X}}
\newcommand{\Y}{\mathbf{Y}}
\newcommand{\Z}{\mathbf{Z}}
\newcommand{\W}{\mathbf{W}}
\newcommand{\R}{\mathbf{R}}
\newcommand{\A}{\mathbf{A}}
\newcommand{\D}{\mathbf{D}}
\newcommand{\B}{\mathbf{B}}
\newcommand{\bb}{\mathbf{b}}
\newcommand{\dd}{\mathbf{d}}
\newcommand{\M}{\mathbf{M}}
\newcommand{\T}{\mathbf{T}}
\newcommand{\F}{\mathbf{F}}
\renewcommand{\S}{\mathbf{S}}
\renewcommand{\H}{\mathbf{H}}
\newcommand{\g}{\mathbf{g}}
\renewcommand{\r}{\mathbf{r}}
\newcommand{\s}{\mathbf{s}}
\renewcommand{\tt}{\mathbf{t}}
\newcommand{\uu}{\mathbf{u}}
\newcommand{\w}{\mathbf{w}}
\newcommand{\x}{\mathbf{x}}
\newcommand{\vv}{\mathbf{v}}
\newcommand{\z}{\mathbf{z}}
\newcommand{\y}{\mathbf{y}}
\newcommand{\RR}{\mathbb{R}}
\newcommand{\ZZ}{\mathbb{Z}}
\newcommand{\NN}{\mathbb{N}}
\newcommand{\XX}{\mathbb{X}}
\newcommand{\WW}{\mathbb{W}}
\newcommand{\II}{\mathbb{I}}
\newcommand{\LL}{\mathcal{L}}

\newcommand{\BBeta}{\boldsymbol{\beta}}
\newcommand{\btheta}{\boldsymbol{\theta}}
\newcommand{\Alpha}{\boldsymbol{\alpha}}
\newcommand{\bTheta}{\boldsymbol{\Theta}}
\newcommand{\bgamma}{\boldsymbol{\gamma}}
\newcommand{\bpi}{\boldsymbol{\pi}}
\newcommand{\arrowp}{\stackrel{p}{\rightarrow}}
\newcommand{\arrowd}{\stackrel{d}{\rightarrow}}
\newcommand{\arrowas}{\stackrel{a.s.}{\rightarrow}}
\newcommand{\approxsim}{\stackrel{approx.}{\sim}}
\newcommand{\0}{\mathbf{0}}
\newcommand{\bP}{\mathbf{P}}
\newcommand{\nn}{\nonumber}
\def\S{{\sigma}}
\renewcommand{\qed}{\space $\square$}

\makeatletter
\newcommand{\oset}[3][0ex]{
  \mathrel{\mathop{#3}\limits^{
    \vbox to#1{\kern-2\ex@
    \hbox{$\scriptstyle#2$}\vss}}}}
\makeatother
\newcommand{\inlinearrowp}{\oset{p}{\rightarrow}}
\newcommand{\inlinearrowd}{\oset{d}{\rightarrow}}

\newcommand{\bigvert}{\, \big | \,}
\newcommand{\Bigvert}{\, \Big | \,}
\newcommand{\biggvert}{\, \bigg | \,}
\newcommand{\Biggvert}{\, \Bigg | \,}

\delimitershortfall-1sp

\makeatletter
\newcommand{\listintertext}{\@ifstar\listintertext@\listintertext@@}
\newcommand{\listintertext@}[1]{
  \hspace*{-\@totalleftmargin}#1}
\newcommand{\listintertext@@}[1]{
  \hspace{-\leftmargin}#1}
\makeatother

\allowdisplaybreaks[1]

\setlength{\parindent}{0pt}
\setlength{\parskip}{6pt}

\addtolength{\oddsidemargin}{-.25in}
	\addtolength{\evensidemargin}{-.25in}
	\addtolength{\textwidth}{.5in}
\addtolength{\textheight}{.2in}
\addtolength{\topmargin}{-.1in}

\makeatletter
\renewcommand*{\ext@figure}{lot}
\let\c@figure\c@table
\let\ftype@figure\ftype@table
\let\listoftableandfigures\listoftables
\renewcommand*\listtablename{Tables and Figures}
\makeatother

\makeatletter
\newcommand{\setword}[2]{
  \phantomsection
  #1\def\@currentlabel{\unexpanded{#1}}\label{#2}
}
\makeatother

\section{Proof of Proposition 4(a)}
Suppose that we have an RCT where SUTVA holds. Then $\hat \mu_{1,R}$ is finite-sample unbiased for $\mu_{1}$.

\begin{proof}

Much of the proof follows the exposition of \citet{aronowmiller}, Theorem 6.2.6.

%

Begin by decomposing into three terms:
\begin{align}
 \E[ \hat \mu_{1,R}] & = \E\left[n^{-1} \sum_{i=1}^n \left\{ \frac{(Y_i -  g(W_i, \hat \beta_i))X_i}{\pi_i(W_i)} + \sum_{i=1}^n g(W_i, \hat \beta_i) \right\}\right] \\
 & = \E\left[\frac{Y_i X_i}{\pi_i(W_i)}\right] -  \E \left[\frac{g(W_i, \hat \beta_i)X_i}{\pi_i(W_i)}\right] + \E \left[g(W_i, \hat \beta_i)\right],
\end{align}
By linearity of expectations and unbiasedness of the sample mean.

We begin by examining the first term ($\E\left[\frac{Y_i X_i}{\pi_i(W_i)}\right]$), and first derive the analogous conditional expectation with respect to $W_i$. By definition,
\begin{align}
    \E\left[\frac{Y_i X_i}{\pi(W_i)} | W_i \right] =\E\left[Y_i \cdot \frac{X_i}{\Pr \left( X_i = 1 \vert W_i\right)} | W_i \right]
\end{align}
And by the potential outcomes model, 
\begin{align*}
    \E\bigg[Y_i \cdot\frac{X_i}{\Pr(X_i = 1 \vert W_i)} \biggvert W_i \bigg] = \E\bigg[Y_i(1) \cdot \frac{X_i}{\Pr \left(X_i = 1 \vert W_i\right)} \biggvert W_i \bigg].
\end{align*}
Then since ignorability holds, $Y_i(1) \independent X_i \vert W_i$, so:
\begin{align*}
\E\bigg[Y_i(1) \cdot \frac{X_i}{\Pr \left([X_i = 1 \vert W_i\right)} \biggvert W_i \bigg]
& = \E\left[Y_i(1) \bigvert W_i \right] \cdot \E\bigg[\frac{X_i}{\Pr(X_i = 1 \vert W_i)} \biggvert W_i \bigg] \\
& = \E\left[Y_i(1) \bigvert W_i \right] \cdot \frac{\E[X_i \vert W_i]}{\Pr(X_i = 1 \vert W_i)} \\
& = \E\left[Y_i(1) \bigvert W_i \right] \cdot \frac{\Pr(X_i = 1 \vert W_i)}{\Pr(X_i = 1 \vert W_i)} \\
& = \E\left[Y_i(1) \bigvert W_i \right].
\end{align*}
Thus, 
$$
\E\bigg[\frac{Y_i X_i}{\pi(W_i)} \biggvert W_i \bigg] = \E\left[Y_i(1) \bigvert W_i \right].
$$
So by the Law of Iterated Expectations,
$$
\mu_1 = \E\big[Y_i(1)\big] = \E\Big[\E\left[Y_i(1) \bigvert W_i\right]\Big] = $$
$$\E\Bigg[\E\bigg[\frac{Y_i X_i}{\pi(W_i)} \biggvert W_i \bigg]\Bigg] = \E\bigg[\frac{Y_i X_i}{\pi(W_i)}\bigg].
$$

The second term ($-\E \left[\frac{g(W_i, \hat \beta_i)X_i}{\pi_i(W_i)}\right]$) obeys the same logic.  Since $\hat\beta_i$ is selected with a sample splitting procedure, we have $\hat \beta_i \independent X_i | W_i$:
\begin{align*}
\E \left[-\frac{g(W_i, \hat \beta_i)X_i}{\pi_i(W_i)} \biggvert W_i \right] & = -\E \left[g(W_i, \hat \beta_i) \biggvert W_i \right] \E \left[\frac{ X_i}{\pi_i(W_i)} \biggvert W_i \right] \\
& = -\E[g(W_i,\hat \beta_i) | W_i].
\end{align*}
Applying the law of iterated expectations, we have $$
\E \left[-\frac{g(W_i, \hat \beta_i)X_i}{\pi_i(W_i)} \biggvert W_i \right] = -\E[g(W_i,\hat \beta_i)]
$$
which cancels with the third term ($\E[g(W_i,\hat \beta_i)]$), leaving:
$ \E[ \hat \mu_{1,R}] = \mu_1.$
\end{proof}

\newpage

\section{Additional Figures}

\begin{figure}[H]
    \centering
    \includegraphics[width=0.39\textwidth]{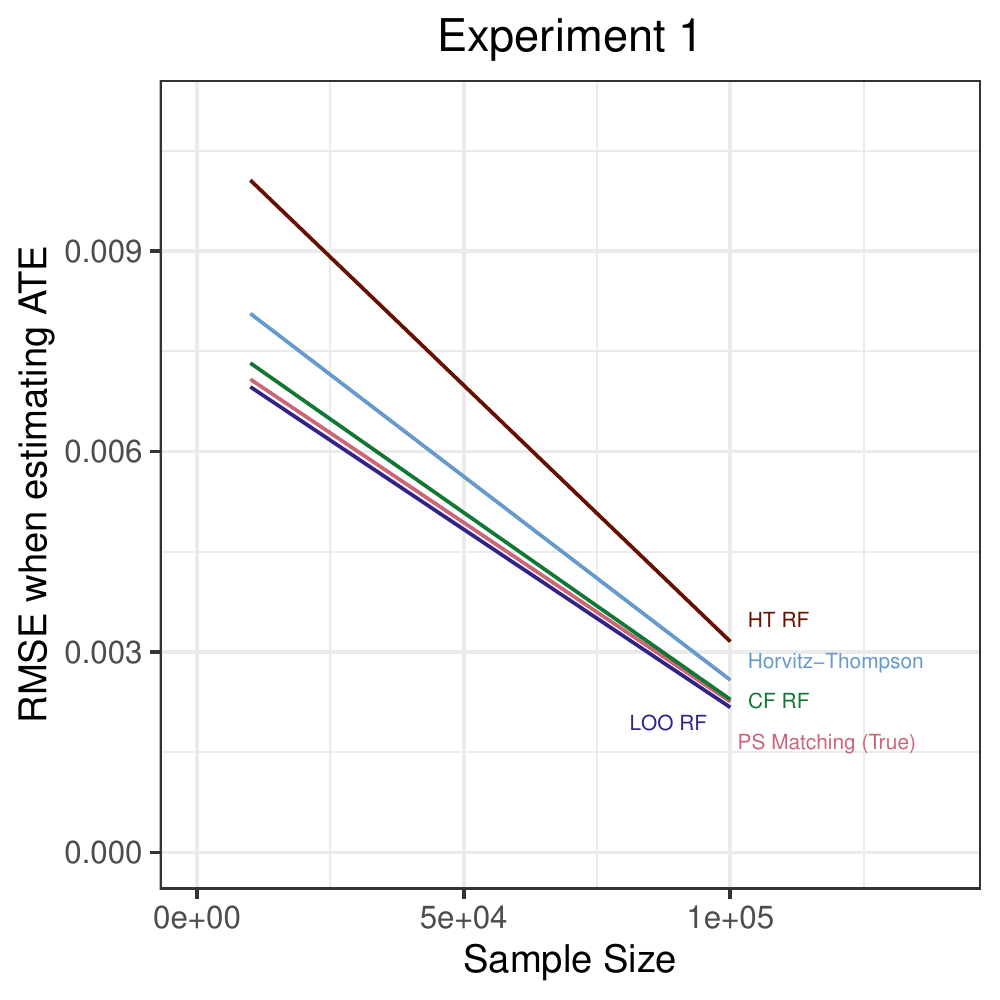}
    \includegraphics[width=0.39\textwidth]{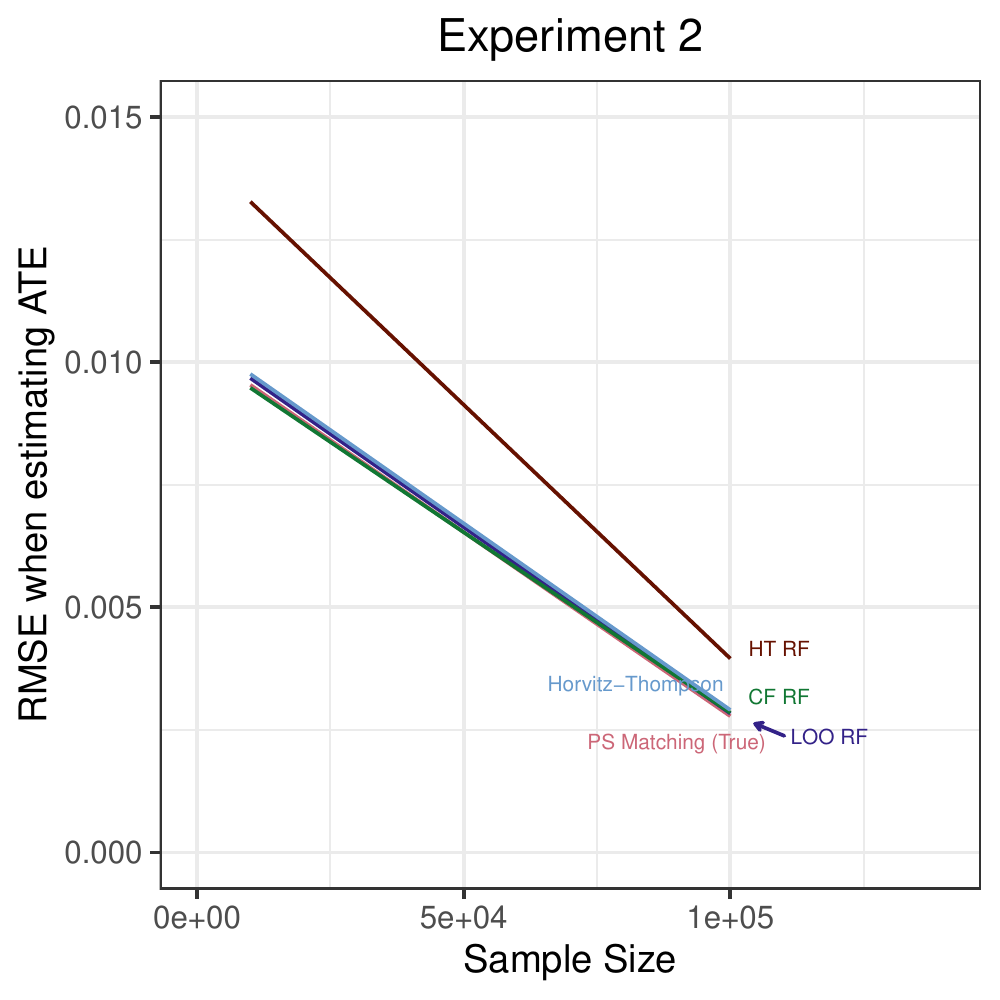}
    \includegraphics[width=0.39\textwidth]{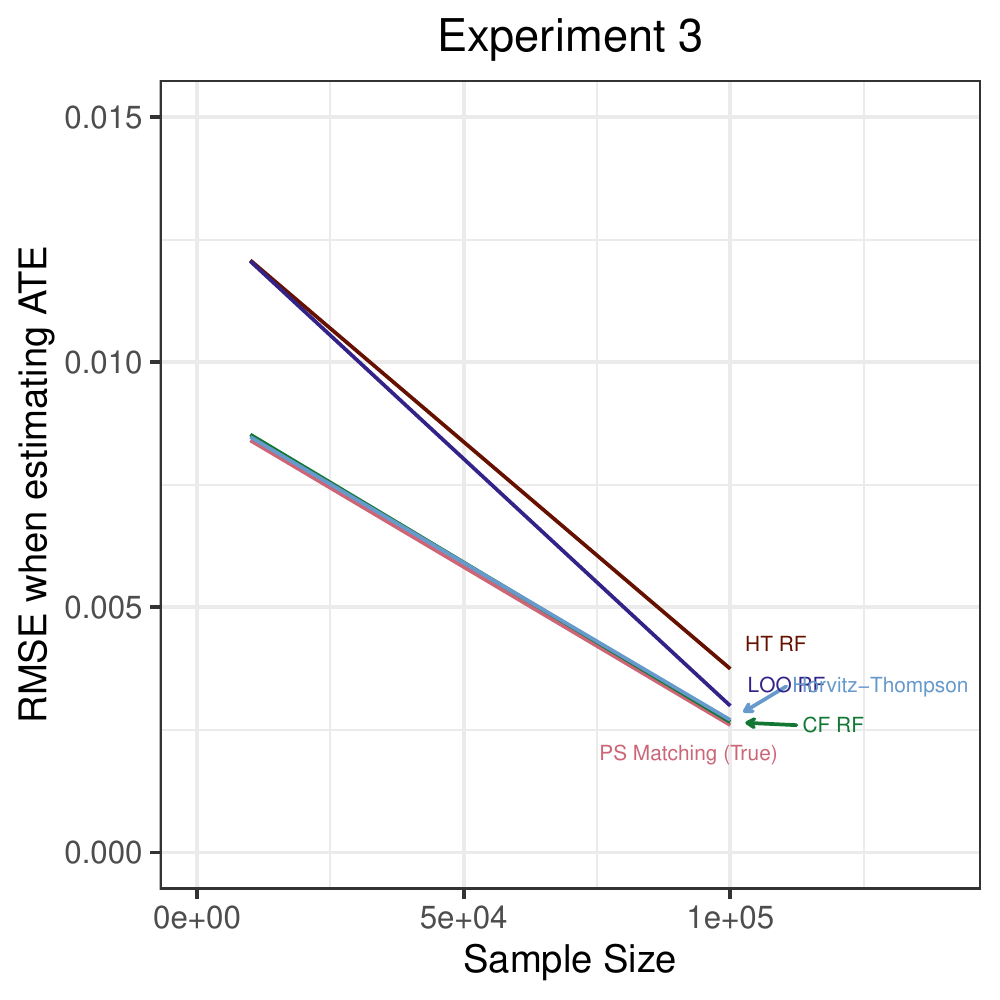}
    \caption{The RMSE of the estimators which use the true propensity score when estimating the average treatment effect for a range of $n$ in the experimental setups outlined in Section \ref{sim_setups}}
    \label{fig:rmse_zoomed}
\end{figure}

\begin{figure}[H]
\centering
    \includegraphics[width=0.45\textwidth]{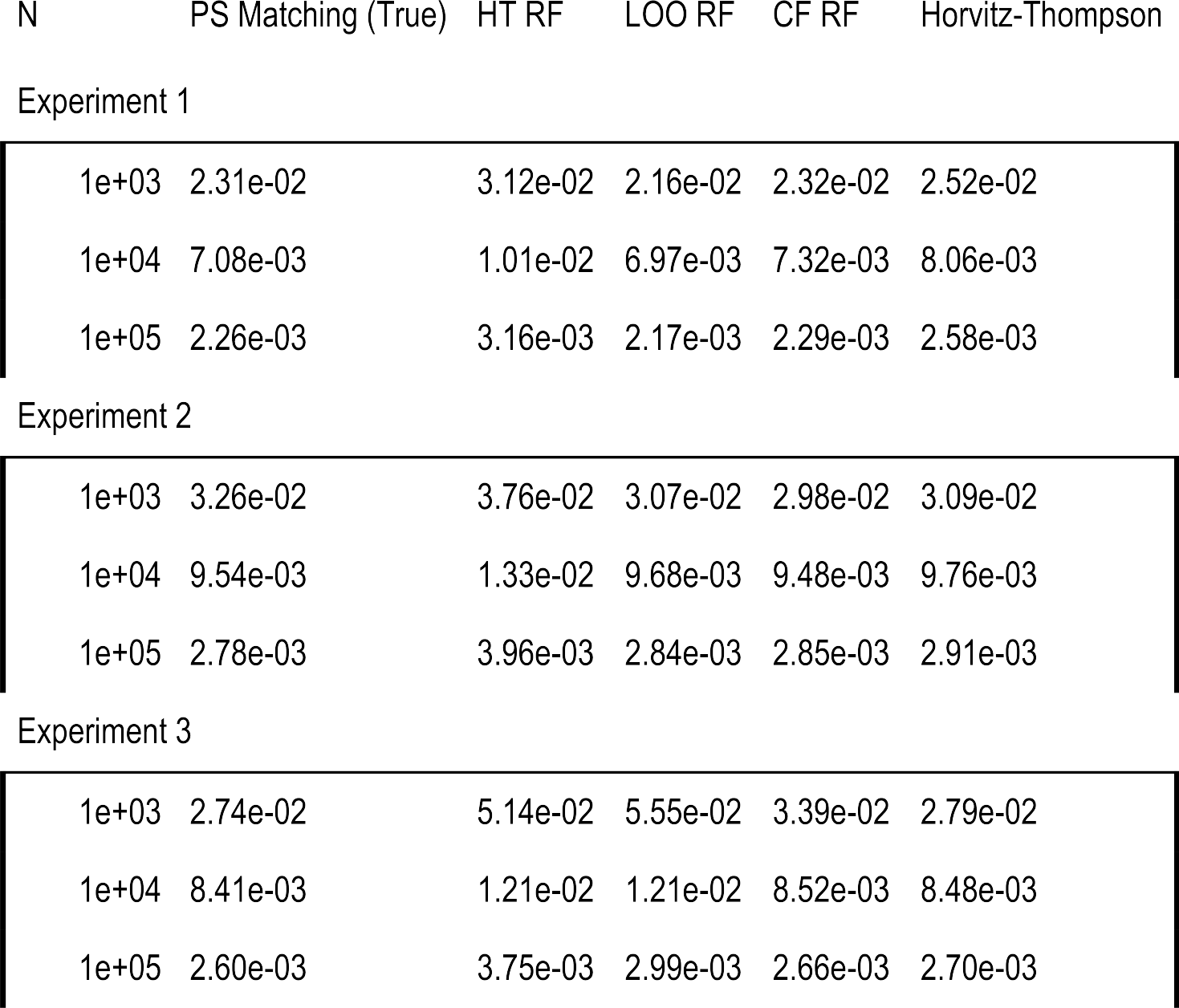}
    \caption{The RMSE of all estimators which use the true propensity score when estimating the average treatment effect for a range of $n$ in the experimental setups outlined in Section \ref{sim_setups}}
    \label{fig:rmse_table}
\end{figure}

\end{document}